\documentclass[english,aps,prstper,reprint,showpacs,titlepage,longbibliography,floatfix]{revtex4-2}   

\usepackage[T1]{fontenc}	
\usepackage[latin9]{inputenc}	
\usepackage{geometry}		
\usepackage{graphicx}
\usepackage[above,below]{placeins}	
\usepackage{times}

\usepackage{hyperref}  
\hypersetup{colorlinks=true,urlcolor=blue,citecolor=blue,linkcolor=blue}   
\usepackage{enumitem}          
\setlist{nosep}                 

\begin{document}

\begin{titlepage}

  \title{Evaluation of a deliberate-practice informed supplemental intervention in graduate Quantum Mechanics}

  \author{Michael E. Robbins, Guillaume M. Laurent}
  \affiliation{Department of Physics, Auburn University, 380 Duncan Drive, Auburn, AL, USA 36849} 
  \author{Eric W. Burkholder}
  \affiliation{Department of Physics, Auburn University, 380 Duncan Drive, Auburn, AL, USA 36849}
  \affiliation{Department of Chemical Engineering, Auburn University, 222 Foy Union Circle, Auburn, AL, USA 36849} 


  \begin{abstract}
Despite the prevalence of physics education research literature related to problem solving, recent studies have illustrated that opportunities for ``authentic'' problem solving---conceptualized as making decisions with limited information using one's physics knowledge---are limited at both the graduate and undergraduate levels in physics curricula. Building on one of these studies, we designed a supplemental intervention for a graduate-level quantum mechanics course which scaffolded the practice of making some of these critical decisions using the conceptual framework of deliberate practice. Despite similar incentive structures as prior interventions focused on conceptual understanding in similar contexts, we did not measure any statistically significant improvement in students' problem solving skills following our intervention, though faculty members involved with the next course and written qualifying exams indicated the students showed better-than-usual conceptual understanding. We explore a number of potential explanations for this disconnect and suggest future avenues of research in this area.\clearpage
  \end{abstract}

  \maketitle
\end{titlepage}

\section{Introduction}
Problem solving is an essential element of physics in both practice and coursework. Consequently, problem solving has been the subject of extensive research in physics education (PER) that spans decades, ranging from clinical observations of student problem solving tendencies \cite{Rahayu_2018} to the development and evaluation of problem solving teaching strategies \cite{Mestre2011,Robbins2025b}. (For comprehensive reviews of this body of literature at the undergraduate level, see \cite{Docktor2014,KuoBook}.) Despite this large body of research, a recent study suggested that opportunities for authentic problem solving remain limited in undergraduate physics curricula \cite{Montgomery2023}. Additionally, there is relatively little research on problem solving and related teaching strategies at the graduate level \cite{Leak2017}, despite reports from recent Ph.D. graduates that they solve technical problems more often than they draw on particular physics content knowledge \cite{AIPSkills}, and employers report graduates often fall short of their expectations in this area \cite{symonds2018}. We thus aim to extend prior PER literature on teaching and learning at the graduate level in the context of quantum mechanics \cite{Singh2008,Zhu2012,Porter2019, Porter2020}.

In this study, we adapt a framework which characterizes authentic problem solving as a process of making 29 different decisions \cite{Price2021}, including deciding on appropriate assumptions and simplifications, determining whether a solution to a problem makes sense, and narrowing the scope of a problem in order to make measured progress. In our prior studies of problem solving in graduate physics curricula, we found that few of the 29 decisions identified in Ref. \cite {Price2021} were explicitly practiced in graduate physics assessments, despite instructors suggesting that graduate students should be able to execute a large number of the decisions by the end of their coursework \cite{Robbins}. We then designed an assessment to measure these decision-making skills \cite{Robbins2025a} based on a template outlined in Ref. \cite{Price2021}, which confirmed that graduate students' decision-making capabilities were limited following their coursework.

In this study, we developed a deliberate-practice intervention to supplement their graduate Quantum Mechanics (QM) coursework. Deliberate practice is a structured, intensive activity with an explicit goal of improving performance  by (1) clearly defining the skills to be practiced, (2) providing timely feedback to students, and (3) allowing students the opportunity to incorporate that feedback \cite{Ericsson1993}, and (4) rewarding effort and progress as opposed to performance on summative assessments \cite{Schwartz2016}. Importantly, the skills practiced must be appropriately tailored to the students' prior knowledge and specific difficulties. Given others' success using supplemental instruction to support students' development of conceptual understanding in graduate QM \cite{Porter2020}, we hypothesized that we may be able to similarly support the development of students' decision-making skills in this same domain.

\section{Methods}

In this study, we developed a set of supplemental assignments for students to practice a subset of the decisions identified in \cite{Robbins}. These assignments were used in a first-semester graduate QM course at a large research university in 2024, and the effectiveness of this supplemental intervention was assessed using a previously-developed assessment of problem solving in QM in comparison with a quasi-experimental control group from 2023 \cite{Robbins2025a}. The assessment was used as a post-test only as our pilot studies suggested that pre/post measurements with this instrument might be biased by students' memories of the assessment over the time span of a single semester. Thus, we instead used a previously-developed QM concept inventory (CI) test \cite{Mason2010} to account for potential differences in the two groups' prior knowledge of QM. We chose to use this specific CI because discipline-specific decision-making requires content knowledge in the discipline \cite{Wieman2019}. 

\subsection{Assignments}
The authors worked together to create a set of student learning outcomes (SLOs) for the course. At the start of each unit, the second author (instructor) created a list of SLOs which was edited by the first author to use more specific and measurable verbs (e.g. ``apply'' instead of ``understand''); all changes were discussed and agreed upon. The first author then developed nine assignments (given in non-exam weeks) which targeted a subset of content-related SLOs and specific decisions. The assessments followed a ``TILT-ed'' design \cite{TILT} which listed the SLOs to be practiced by the assignment and gave an example of the decisions to be practiced. In contrast to the course homework, the assignments required few or no calculations, and instead focused on particular decisions (summarized in Table \ref{outcomes}) including: what are the important features of the problem (\#4), what are related problems (\#7) or potential solutions (\#8), whether the result matches expectations (\#20) and if it makes sense (\#26), what simplifications are appropriate (\#10), how to decompose the problem and make it more tractable (\#11), and whether previous assumptions are still appropriate (\#23). The decisions were introduced gradually and revisited in future assignments to allow students spaced, repeated practice \cite{Karpicke2011}; see the textbox below for an example assignment.

\begin{table}[h]
  \caption{List of course content and objectives. $^\star$Indicates the decision-making outcome was introduced for the first time.\label{outcomes}}
  \begin{ruledtabular}
    \begin{tabular}{lll}
      \textbf{Week}
      & \textbf{Course Topic} 
      & \textbf {Decisions Involved \cite{Price2021}} \\ 
      \hline
     3 & Mathematical Tools of QM & $26^\star$ \\
     5 & Postulates of QM & $7^\star,8^\star$ \\
     6 & Postulates of QM & $20^\star,26$ \\
     7 & 1-D Problems & $10^\star$ \\
     10 & 1-D Problems & $11^\star$ \\
     11 & Angular Momentum & $4^\star, 8 , 11$ \\
     12 & Angular Momentum & $7, 20, 23^\star, 26$ \\
     14 & 3-D Problems & $8, 26$ \\
     15 & 3-D Problems & $ 11$ \\
    \end{tabular}
  \end{ruledtabular}
\end{table}

\subsection{Intervention and Course Context}
Data were collected from a primarily lecture-based, first-semester graduate QM course at a large research university in two consecutive offerings of the course (Spring 2023 and Spring 2024); both offerings were taught by the second author. In the first offering (``Control''), the instructor taught the course as he had for the previous two years.  All 15 students invited in the control group agreed to participate and completed the pretest (CI); 10 completed the post-test. In the intervention course, 18 students were invited, and 16 agreed to participate in the study and completed the pretest (9 completed the post-test).  The pretest was completed in class during the first week of the course and the post-test was completed outside of class within one week of the final exam. No demographic information was taken to preserve participant confidentiality. 

\vspace{2mm}

\newcommand{\SubItem}[1]{
    {\setlength\itemindent{15pt} \item[-] #1}
}
\noindent\fbox{%
    \parbox{\columnwidth}{%
            \textbf{Quantum Learning Objectives} \\
            A student should be able to:
            \begin{itemize}
            \item recall that no pair of angular momentum components commute.
            \item recall that $J^2$ commutes with its components.
            \item...
            \end{itemize}
            \textbf{Problem Solving Learning Objectives}\\
            A student should be able to:
            \begin{itemize}
            \item Identify the important underlying features or concepts that apply. This could include
                \SubItem{Which available information is relevant to solving and why?}
                \SubItem{Create/find a suitable abstract representation of core ideas and information (i.e., an equation).}
            \end{itemize}
            \textbf{Problems}\\
        	A system has the wave function $\psi$. $J^2$ is measured to be $2 \hbar^2$, and $J_z$ is measured to be $-\hbar$. We want to determine which state of $J_z$ leads to the lowest uncertainty of$J_x: \Delta J_x$.
            \begin{itemize}
            \item Identify  all the necessary information to determine this expected value. Explain why it's necessary.
            \item Without  calculating, which state would you predict. Explain.
            \item Describe  which steps you would take to find this expected value. (Do not solve)            \end{itemize}

    }%
}

\vspace{2mm}

We deployed the supplemental assessments as the ``Intervention'' in 2024, offering extra credit for completing the assessments (see Table \ref{grading}). In the Control, in-class exams were 75\% of the grade and the final exam was 25\% of the grade. The instructor provided homework for practice which was not graded. In the Intervention, the assignments were offered as a supplement to the non-graded homework. The assignments were distributed digitally once a week in non-exam weeks, beginning the third week of class. Submissions were graded by the first author using a coarse rubric: no attempt, 0\%; major mistakes 50\%; moderate mistakes 80\%; minor mistakes, 100\%; the median score was 100\% Detailed feedback was provided, highlighting exceptional aspects of a response or areas to improve. Completing the 9 weekly assignments and post-test (counted as two weekly assignments) offered students the chance to earn up to a 5\% bonus on their overall course grade. In addition, each in-class exam included a bonus problem (scored by the first author) related to the decisions practiced in that unit; students also received a 5\% bonus on their exam grades for completing these problems. 

\begin{table}[htbp]
  \caption{Grading structor of the control and intervention courses. $^\star$ Indicates bonus points, allowing for a score greater than 100\%\label{grading}}
  \begin{ruledtabular}
    \begin{tabular}{lll}
      \textbf{Category}
      & \textbf{Control} 
      & \textbf {Intervention} \\
      & \textbf{Spring 2023} & \textbf{Spring 2024}\\
      \hline
     In-Class Exams & $75\%$ & $75\%$ \\
     Final Exam & $25\%$ & $25\%$ \\
     Homework & & $5\%^\star$ 
    \end{tabular}
  \end{ruledtabular}
\end{table}

\section{Results}
\begin{figure}
  \includegraphics[width=0.95\linewidth]{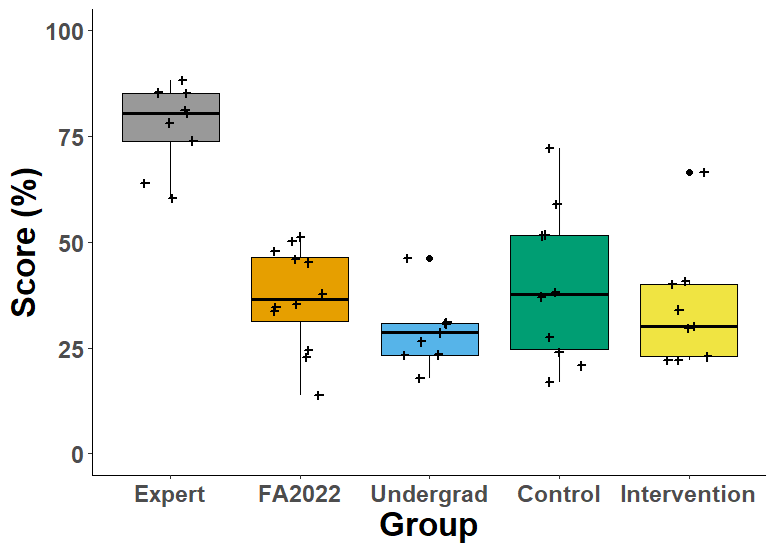}
  \caption{Box-and-whisker plot of scores (as a percentage) on the QM assessment. The expert, FA2022 (graduate student) and Undergrad groups were scores from the initial validation of the QM assessment \cite{Robbins2025a}, to provide a reference for the Control (Spring 2023 post-test) and Intervention (Spring 2024 post-test) groups. The group medians are indicated by solid black lines and the lower and upper bounds of each box are the 25th and 75th percentiles, respectively. The whiskers indicate the 10th and 90th percentiles, and any large black dots indicate outliers, while the smaller cross-signs are the individual data points. \label{fig:AssessmentResults}}
\end{figure}

The median post-test QM assessment score was 37.5\% for the control group and 34.2\% for the intervention group. These scores were also similar to the scores of graduate students in the assessment validation study (see Figure \ref{fig:AssessmentResults}). A Mann-Whitney test confirms that these results are not statistically distinguishable at the $\alpha = 0.05$ level. Similarly, the pretest scores are not statistically distinguishable at the same level (Control median: 21.0 \%, Intervention median: 19.5 \%). 

\begin{figure*}
  \includegraphics[width=0.8\linewidth]{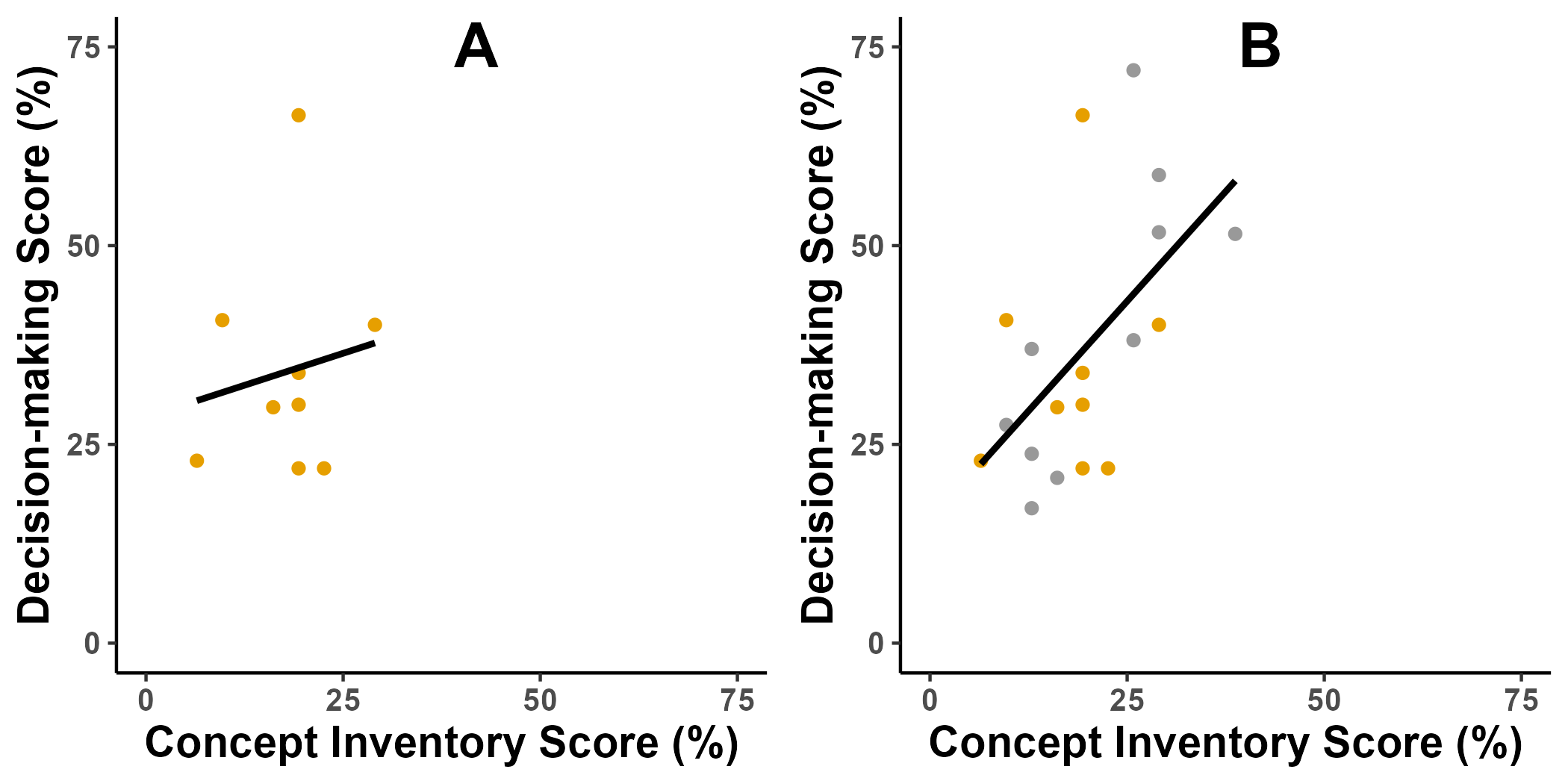}
  \caption{QM assessment scores versus concept inventory scores for (A) the intervention group and (B) the control and intervention groups combined. In both plots the control group is grey and the intervention group is orange. Both axes represent scores scaled to be from 0 to 1. \label{fig:AssessmentCorrelations}}
\end{figure*}

The raw scores suggest that the two groups had equivalent levels of incoming conceptual knowledge of QM and equivalent post-test scores on the QM assessment. Given that the development of the QM assessment identified a correlation between the CI and QM assessment as a proxy of validity, this would suggest that the intervention had no measurable effect and that this could not be attributed to lower levels of prior knowledge in the Intervention group. However, a paired Spearman test indicated that the correlation between CI scores and post-test scores in the Intervention group were almost non-existent ($\rho = 0.02$), as shown in Figure \ref{fig:AssessmentCorrelations}. Curiously, this suggests that participants scoring higher on the CI test did not perform any better on the post-test than those scoring lower on the CI. The results differ from the strong correlation found in the control group ($\rho = 0.82$). One interpretation may be the intervention was detrimental to those with the highest score on the CI test. However, these results were likely due to statistical fluctuations. When conducting a paired Spearman test for the control and intervention groups combined, the results suggest a moderate correlation ($\rho = 0.54$). The lack of correlation in the intervention group may be explained by the limited range of the data and low number of samples. Indeed, panel B in Figure \ref{fig:AssessmentCorrelations} suggests that there are many higher CI scores in the control group compared with the intervention group which are also associated with higher assessment scores.

\begin{figure}
  \includegraphics[width=0.8\linewidth]{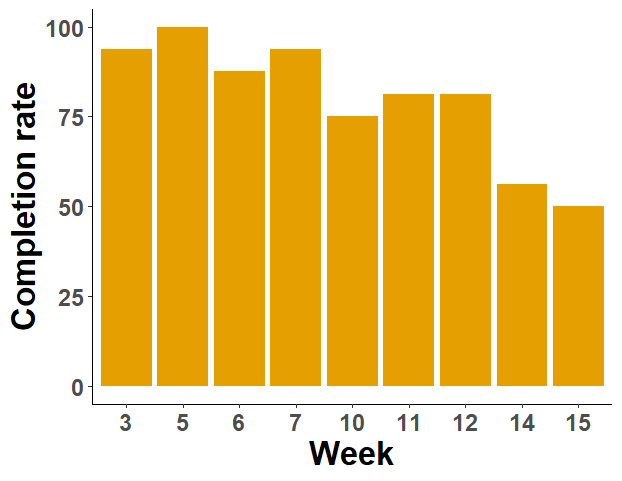}
  \caption{The number of completed assignments out of 16 students for each assignment, ordered by week of the semester.\label{fig:AssessmentCompletion}}
\end{figure}

Because the supplemental assignments were optional, we sought to determine if the number of assignments completed was correlated with students' performance on the post-test. As shown in Figure \ref{fig:AssessmentCompletion}, the assignment completion rate decreased as the semester progressed. However, of the 9 students who completed the post-test, 7 completed all assignments, with an overall average completion rate of 84\% and a median completion rate of 100\%. Of the 16 students who participated, 8 completed all assignments, with an overall average completion rate of 80\% and a median completion rate of 85\%. This indicates that the students in our post-test dataset were primarily students who were completing all of the assignments. Therefore, we cannot conclude that the decrease in completion explains the measured effect of the intervention.

\section{Discussion \& Conclusions}
In this study, we developed an intervention comprised of assignments which targeted a small set of decision-making skills through deliberate practice. Using a quasi-experimental design, we were unable to measure a significant impact of the intervention on students' decision-making skills as measured by an assessment we had designed for this purpose. This null result does not seem to be tied to students' participation in the supplemental intervention or differences in the two groups' incoming knowledge of QM.

This null result was surprising to the instructor, who felt that the students showed a deeper conceptual understanding of the material on the second exam. For example, students seemed to engage in predicting potential solutions to a problem (decision 8) before beginning their calculations on exams--something he had never seen before in teaching this class and which was explicitly practiced on supplemental assignment 5 right before this exam. Independent parties also reported similar patterns of deeper conceptual understanding when the Intervention cohort completed their written qualifying exams later that year. This was also surprising as these exams are typically focused on mathematical fluency and warrants further analysis of evidence of decision-making in students' solutions to the qualifying exams. 

Though the point incentive for students to participate in this intervention were minimal, our initial investigation of students' completion of and effort toward the supplemental assignments did not point to this being an obvious cause for the null results. Indeed, despite the time constraints faced by many first-year graduate students, each assignment had a majority of participants complete it, and the total time required of participants was estimated to be 11 hours over the semester. This means, however, that the overall time dedicated to deliberate practice is relatively low compared to the overall course; a typical three credit-hour course assumes an average of 150 hours over a sixteen-week semester. Furthermore, this limited time also meant that we were not able to address all of the decisions probed by the assessment with sufficient time for feedback and additional practice. Similarly, as seen with the example above, the problems themselves were typically not situated in authentic contexts, such as those a practicing physicist might encounter. Though the underlying skills practiced might be the same, students may also require targeted practice identifying which decisions are needed or relevant when approaching a more realistic problem. 

The intervention described in this study drew upon well-established frameworks for learning and mastering new skills and requires little to no course modifications to adopt. However, this approach did not result in any measurable changes in student problem-solving skills. A logical next step would be to enhance this intervention to further support the development of decision making skills. Other studies of teaching these types of skills have typically thoroughly integrated decision-making practice throughout all elements of the course, rather than as an additional piece \cite{Burkholder2022,Burkholder2021}. We initially adopted the supplemental approach given its success in other contexts in supporting students' conceptual understanding of QM \cite{Porter2020}.

A more integrated approach could help address potential issues of student motivation because all practice of decision-making becomes crucial to their overall success in the course. Additionally, incorporating decision-making  throughout course problems provides additional time to practice these skills. Rather than a total course redesign, a more modest change may include modifying course problems to have a written response portion. For example, an add-on to existing problems may ask for justification to a solution's correctness or more complex problems may be added which require a solution plan instead of detailed calculations. Additionally, lectures may include intentional explanations of certain decisions, such as why certain assumptions were acceptable for one problem but not another. These proposed enhancements require a larger portion of the finite course time, and a more active adaptation of these methods.

Another open question relates to the way we are measuring these decision-making skills. We adopted for our own specially-designed assessment of decision-making as we had previously found that typical exam problems were calculation-heavy and focused relatively little on higher-order problem-solving skills. While this assessment is better-suited to our ultimate goals, there are many more studies of its validity and reliability which will be required to understand how it is best used. For example, it may simply be that one semester is too short of a time-scale to measure changes in expert-like decision making regardless of interventions. Or, if it is possible, the effect size may be small enough that the sample size included in this study is insufficient to measure this effect. 

Finally, though our study arguably raises more questions than it provides answers, we felt it important to report these results to help avoid publication bias in the PER literature on classroom interventions \cite{Van_Lent2014} and contribute to the broader literature on when and why certain interventions work.

\acknowledgments{This material is based upon work supported by the National Science Foundation under Grant No. DGE-2429155.Any opinions, findings, and conclusions or recommendations expressed in this material are those of the author(s) and do not necessarily reflect the views of the National Science Foundation.}

\bibliography{bibfile} 

\end{document}